\documentclass[a4paper,fleqn]{cas-dc}
\usepackage[authoryear,longnamesfirst]{natbib}
\usepackage{times}
\usepackage{epsfig}
\usepackage{graphicx}
\usepackage{amsmath}
\usepackage{amssymb}
\usepackage{afterpage}
\usepackage[ruled,linesnumbered]{algorithm2e}
\usepackage{bm}
\usepackage{booktabs}
\usepackage[switch]{lineno}
\usepackage[switch]{lineno}
\usepackage{tabularx}
\usepackage{booktabs}
\usepackage{microtype}
\usepackage{subcaption}

\usepackage{algorithm}
\usepackage{algorithmic} 
\usepackage{caption}
\usepackage{multirow} 
\usepackage{balance}
\graphicspath{{./image/}}
\def\tsc#1{\csdef{#1}{\textsc{\lowercase{#1}}\xspace}}
\tsc{WGM}
\tsc{QE}

\begin{document}
\let\WriteBookmarks\relax
\def\floatpagepagefraction{1}
\def\textpagefraction{.001}

\shorttitle{Experimental Study on Surveillance Video-Based Indoor Occupancy Measurement for Occupant-Centric Control}    

\shortauthors{Irfan Qaisar et al.}

\title[mode=title]{Experimental Study on Surveillance Video-Based Indoor Occupancy Measurement for Occupant-Centric Control}  

\author[1]{Irfan Qaisar}[type=editor,style=chinese, orcid=0000-0002-4831-977X]
\fnmark[1]
\ead{irfan21@mails.tsinghua.edu.cn}

\author[2,3]{Kailai Sun}[type=editor,style=chinese,orcid=0000-0003-1648-3409]

\fnmark[1]
\cormark[1]

\ead{skl24@mit.edu}

\author[1]{Qingshan Jia}[type=editor,style=chinese,orcid=0000-0002-4683-7215]
\ead{jiaqs@tsinghua.edu.cn}

\author[1]{Qianchuan Zhao}[type=editor,style=chinese,orcid=0000-0002-7952-5621]
\cormark[1]


\ead{zhaoqc@mail.tsinghua.edu.cn}

\affiliation[1]{organization={Center for Intelligent and Networked Systems, Department of Automation, BNRist, Tsinghua University},
            city={Beijing},
            postcode={100084}, 
            country={China}}


\affiliation[2]{organization={Urban Mobility Lab, Massachusetts Institute of Technology},
            addressline={Cambridge}, 
            state={MA 02139},
            country={USA}}
            
\affiliation[3]{organization={Singapore-MIT Alliance for Research and Technology Centre, Massachusetts Institute of Technology},
            postcode={138602}, 
            country={Singapore}}

\cortext[1]{Corresponding authors.}

\fntext[1]{These authors contributed equally to this work.}


\begin{abstract}
Accurate occupancy information is essential for closed-loop occupant-centric control (OCC) in smart buildings. However, existing vision-based occupancy measurement methods often struggle to provide stable and accurate measurements in real indoor environments, and their implications for downstream HVAC control remain insufficiently studied. To achieve Net Zero emissions by 2050, this paper presents an experimental study of large language models (LLMs)-enhanced vision-based indoor occupancy measurement and its impact on OCC-enabled HVAC operation. Detection-only, tracking-based, and LLM-based refinement pipelines are compared under identical conditions using real surveillance data collected from a research laboratory in China, with frame-level manual ground-truth annotations. Results show that tracking-based methods improve temporal stability over detection-only measurement, while LLM-based refinement further improves occupancy measurement performance and reduces false unoccupied prediction. The best-performing pipeline, YOLOv8+DeepSeek, achieves an accuracy of 0.8824 and an F1-score of 0.9320. This pipeline is then integrated into an HVAC supervisory model predictive control framework in OpenStudio-EnergyPlus. Experimental results demonstrate that the proposed framework can support more efficient OCC operation, achieving a substantial HVAC energy-saving potential of 17.94\%. These findings provide an effective methodology and practical foundation for future research in AI-enhanced smart building operations.
\end{abstract}

\maketitle

\section{INTRODUCTION}
\label{section1}

Buildings represent a major source of global energy consumption and greenhouse gas emissions, accounting for a significant portion of global energy use \cite{GlobalABC2024}. Achieving the net-zero target by 2050 therefore requires substantial reductions in carbon emissions from the building sector \cite{IEA_Buildings_NZE}. Within this context, heating, ventilation, and air-conditioning (HVAC) systems typically constitute the dominant share of operational energy demand, with their contribution commonly reported to be in the range of 40--50\% in commercial and institutional buildings \cite{EnergyHVAC2024}. As modern HVAC systems increasingly operate as closed-loop cyber-physical systems, reliable occupancy information has become an essential feedback input for enabling energy-efficient and demand-responsive control strategies \cite{yuan2024review,huang2024state}. Beyond energy performance, occupancy data also plays a critical role in supporting indoor environmental quality (IEQ), ensuring human health, and enabling safety-related functions such as emergency management and situational awareness \cite{jiang2023pandemic}.

To achieve Net Zero emissions by 2050, occupant-centric control (OCC) aims to use real-time information on occupancy and indoor conditions to adjust HVAC and other building systems according to actual demand \cite{nagy2023ten}. OCC has gained increasing attention because conventional schedule-based operation often ignores real occupancy patterns, which can increase energy use and reduce comfort \cite{PANG2020115727}. IEA EBC Annex 66 emphasizes that occupant presence and behavior strongly influence building energy use and building performance \cite{annex2022definition}. IEA EBC Annex 79 highlights a broader shift from static schedule-based operation toward adaptive, demand-responsive strategies that explicitly account for occupant needs, behavior, and interactions with building systems \cite{annex2024occupant}. Previous studies show that occupant-centric HVAC control can achieve substantial energy savings while maintaining acceptable indoor conditions and comfort, with reported savings of about 10-40\% in building simulation \cite{winkler2020office,qaisar2025experimental}. Among the inputs required by OCC, occupancy is one of the most important because it directly affects when and how building systems should operate \cite{hahn2022information}. If occupancy is estimated inaccurately, HVAC systems may condition vacant spaces or apply setback during actual occupancy \cite{lei2022practical}. This can reduce energy efficiency and thermal comfort, especially because occupants’ thermal states and adaptive behavioral responses are sensitive to temporal variations in indoor environmental conditions \cite{rentala2024application}. The issue is especially important for model predictive control (MPC), where future control actions depend on forecasts of system states and exogenous inputs such as occupancy \cite{yang2021impact,saloux2025theory}. Therefore, accurate and temporally stable occupancy information is essential for effective OCC implementation in practice.

\begin{table}[h]
\centering
\begin{tabular}{|ll|}
\hline 
\multicolumn{2}{|l|}{\textbf{List of Abbreviations}}  \\
CO$_2$  & Carbon Dioxide \\
DL      & Deep Learning \\
DeepSORT    & Deep Simple Online and Realtime Tracking \\
HVAC    & Heating, Ventilation, and Air Conditioning \\
IEQ     & Indoor Environmental Quality \\
LLM     & Large Language Model \\
MAE     & Mean Absolute Error \\
ML      & Machine Learning \\
ID-switch & Identity Switch \\
MOT     & Multi-Object Tracking \\
MPC     & Model Predictive Control \\
OCC     & Occupant-Centric Control \\
PIR     & Passive Infrared \\
RMSE    & Root Mean Squared Error \\
SORT    & Simple Online and Realtime Tracking \\
VLM     & Vision Language Models \\
YOLO    & You Only Look Once \\
\hline
\end{tabular}
\end{table}

Many sensors (e.g., passive infrared (PIR) sensors, CO2 sensors, temperature sensors, humidity sensors, WiFi) have been explored for occupancy measurement and prediction in buildings \cite{chaudhari2024fundamentals,sun2024dmff}. However, these methods often rely on indirect indicators and struggle to provide reliable occupant counts in real indoor environments, especially under dynamic and complex usage conditions \cite{ding2022review,shokrollahi2024passive}. Occupancy prediction methods based on historical data have also been widely studied for building control applications \cite{zheng2026towards}. Yet their performance can degrade when occupancy behavior becomes irregular, event-driven, or highly dynamic, which limits their reliability for real-time control \cite{zhang2022review,li2024systematic}. These limitations motivate the need for more robust and control-relevant occupancy measurement approaches.

Vision-based occupancy measurement has attracted growing attention because cameras provide direct information on indoor occupant presence and count \cite{dino2022vision,zhang2025vision}. Prior studies have explored scene-based counting, head detection, and fusion frameworks to improve estimation performance \cite{xing2023mitp,sun2022fusion}. However, robust deployment in real environments remains challenging due to occlusion, background clutter, head-like objects, limited camera coverage, and temporal instability \cite{amiri2025privacy}. To improve temporal consistency, tracking methods such as SORT, DeepSORT, ByteTrack and MOTIP are widely used to associate detections across frames \cite{sempere2024new,bewley2016simple,wojke2017simple,zhang2022bytetrack,MOTIP}. While these methods can reduce short-term fluctuations \cite{kim2025real}, their effectiveness remains constrained by detection errors under occlusion, crowding, and rapid motion, which may cause identity switches and fragmented trajectories \cite{sun2025towards}.

Traditional vision-based methods still have limited performance in complex indoor environments, especially when detection and tracking outputs are unstable over time \cite{amiri2025privacy,sun2025towards}. Recent large language models (LLMs) and vision-language models (VLMs) have shown strong reasoning abilities \cite{SUN2025128791}, which can support higher-level consistency checking and contextual inference beyond conventional perception pipelines \cite{achiam2023gpt,liu2023visual,lu2024deepseek}. This suggests that LLMs may help improve traditional occupancy measurement methods by refining noisy perceptual outputs. However, such reasoning-based refinement has been only rarely explored for indoor occupancy measurement in buildings and remains largely under-studied in the building energy and HVAC control literature \cite{guo2025deepseek}. In addition, most existing studies focus on improving occupancy accuracy, while relatively little attention has been paid to how occupancy quality affects downstream building control performance \cite{esrafilian2022impact}. More importantly, the energy-saving potential of such approaches still needs to be verified in experimental practice \cite{lin2023performance}.

To bridge these gaps, this study conducts an experimental investigation of surveillance-video-based indoor occupancy measurement for occupant-centric HVAC operation. Representative perception pipelines, including detection-only, tracking-based, and LLM-enhanced approaches, are evaluated under identical conditions using a real indoor surveillance dataset with frame-level manual ground-truth annotations. Their performance is assessed in terms of accuracy and temporal stability. In addition, the best-performing pipeline is integrated into an MPC-based HVAC control framework in OpenStudio and EnergyPlus to examine its impact on building energy use and occupant comfort.

The main contributions of this study are as follows:
\begin{itemize}
    \item An experimental comparison of surveillance-video-based indoor occupancy measurement pipelines, including detection-only, tracking-based, and LLM-based refinement approaches.
    \item A comprehensive evaluation using a real indoor surveillance dataset collected from a research laboratory at Tsinghua University, Beijing, with frame-level manual ground-truth annotations.
    \item A demonstration that tracking improves temporal stability, while LLM-based refinement further enhances measurement performance and reduces false unoccupied predictions; the best-performing pipeline, YOLOv8+DeepSeek, achieves an accuracy of 0.8824 and an F1-score of 0.9320.
    \item An integration of the best-performing pipeline into an MPC-based HVAC supervisory control framework in OpenStudio--EnergyPlus, achieving an HVAC energy-saving potential of 17.94\% for more efficient OCC operation.
\end{itemize}

\section{LITERATURE REVIEW}
\label{section2}

This section surveys existing research on indoor occupancy measurement for building energy and control applications, with emphasis on sensing modalities, vision-based approaches, multi-object tracking techniques, and emerging vision--language methods. 

\subsection{Occupancy information for occupant-centric control}

OCC has gained increasing attention in building control research because occupant presence and density directly affect space conditioning demand, ventilation requirements, and perceived comfort. Recent review studies characterize OCC as a shift away from static, schedule-based HVAC operation toward adaptive, demand-responsive strategies that rely on occupancy and contextual information as key control inputs \cite{yuan2024review,huang2024state,hahn2022information}. In HVAC operation, occupancy influences sensible and latent heat loads, CO$_2$ generation, and the feasibility of zone-level setpoints \cite{liu2025multi}. As a result, accurate occupancy information is widely regarded as a foundational input for closed-loop control architectures and supervisory control strategies, including MPC \cite{zhou2025evaluating}. Empirical and prototype-based investigations further demonstrate that access to real-time occupancy information can lead to measurable energy savings relative to fixed scheduling strategies, while maintaining or improving indoor comfort, particularly in spaces with intermittent usage patterns \cite{pang2025longitudinal}.

Despite these demonstrated benefits, several reviews emphasize that occupant count remains both one of the most influential and one of the most uncertain variables in practical OCC deployments \cite{yuan2024review,huang2024state}. Inaccurate occupancy measurement can propagate through ventilation and thermal control loops, resulting in over-ventilation or under-ventilation, unstable system behavior, or unnecessary conditioning of unoccupied zones \cite{hahn2022information,jiang2023pandemic}. These challenges highlight the importance of occupancy measurement pipelines that provide not only accurate estimates but also temporally stable signals suitable for real-time control.

\subsection{Non-vision occupancy sensing and occupancy prediction}

A wide range of non-vision sensing modalities has been explored for occupancy inference, including passive infrared (PIR) sensors, ultrasonic and door sensors, environmental sensing such as CO$_2$, plug-load measurements, and wireless signals derived from Wi-Fi, Bluetooth, and cellular networks \cite{chaudhari2024fundamentals,song2025impact}. These approaches are often attractive due to their relatively low cost and perceived privacy advantages; however, they typically rely on indirect indicators of occupancy \cite{wu2024novel}. As a result, their performance may be affected by sensing latency (e.g., CO$_2$ mixing dynamics), limited spatial resolution, or reduced robustness under changing occupant behavior and ventilation conditions \cite{shokrollahi2024passive,khan2024occupancy}. Audio-based sensing and multimodal fusion techniques have also been investigated, yet their robustness can be constrained by acoustic variability and background noise, and their practical adoption may be influenced by privacy and data governance considerations \cite{chaudhari2024fundamentals,sun2024dmff}.

Occupancy prediction methods based on statistical modeling, classical ML, and DL have been proposed to forecast future occupancy states in support of proactive HVAC control. However, recent reviews indicate that real-world occupancy behavior is frequently non-stationary and event-driven, with irregular meetings, schedule deviations, and external disruptions that limit the generalization of prediction models across time and building contexts \cite{zhang2022review,li2024systematic,sun2020review}. This limitation is particularly relevant for experimental benchmarking studies such as the present work, where the focus is on measuring occupancy reliably from real video data rather than predicting future states from historical patterns.

\subsection{Vision-based occupancy measurement in buildings}

Computer vision has been widely recognised as a direct approach for estimating occupant presence, count and activity. It enables explicit observation of people rather than relying on indirect proxies \cite{choi2021review}. The feasibility of vision-based occupancy measurement has increased because surveillance cameras are already deployed in many commercial and institutional buildings, and advances in edge computing enable local inference with reduced reliance on cloud processing \cite{choi2021review,zhang2024deep}. Recent reviews report substantial progress in DL based detection and counting methods, while also noting sensitivity to camera viewpoint, occlusion, illumination variability, and crowd density \cite{chaudhari2024fundamentals,zhou2025evaluating}.

From a computer vision perspective, advances in object detection have been driven by improved model architectures and large-scale datasets. The YOLO family remains widely adopted for real-time person detection, including YOLOv4 and YOLOv7 \cite{bochkovskiy2020yolov4,wang2023yolov7}. More recently, YOLOv8 has gained broad adoption as a high-performance detector optimized for efficient deployment across edge and server platforms \cite{ultralytics2023yolov8,afifah2026yolov8}. Nevertheless, building-domain studies consistently report that detection-only, frame-level pipelines can produce unstable occupancy estimates due to missed detections and false positives under occlusion, reflections, motion blur, or partial visibility \cite{choi2021review,zhang2024deep}. These limitations motivate the use of temporal aggregation, tracking, or post-processing strategies rather than relying solely on instantaneous detection outputs.

\subsection{Multi-object tracking for identity-consistent counting}

Multi-object tracking (MOT) improves temporal consistency by associating detections across frames, enabling identity-consistent counting and reducing frame-to-frame fluctuations in estimated occupancy \cite{sun2025towards}. Early tracking-by-detection approaches such as simple online and realtime tracking (SORT) achieved real-time performance using Kalman filtering and Hungarian assignment \cite{bewley2016simple}. DeepSORT extended this framework by incorporating appearance embeddings to mitigate identity switches under moderate occlusion \cite{wojke2017simple}. More recently, ByteTrack improved robustness by associating both high- and low-confidence detections, and it is commonly regarded as a strong baseline in contemporary MOT evaluations \cite{zhang2022bytetrack}.

Tracking performance is commonly evaluated using metrics that capture both detection quality and association accuracy, including CLEAR MOT metrics (MOTA, MOTP), identity-based metrics such as IDF1, and the more recent HOTA metric \cite{MOTrack,luiten2021hota}. From the perspective of occupancy measurement, identity switches and track fragmentation are particularly relevant, as they can distort unique-ID-based counting even when frame-level detection accuracy appears acceptable \cite{wojke2017simple,zhang2022bytetrack}. These considerations motivate the inclusion of ID-switch and fragmentation metrics alongside conventional occupancy error measures in this study.

\subsection{Vision-language and LLM-based refinement of perception outputs}

LLMs and vision-language models (VLMs) have recently demonstrated strong capability in interpreting structured perception outputs and performing higher-level consistency checking and contextual reasoning \cite{achiam2023gpt,hurst2024gpt}. Open research models such as Flamingo, BLIP-2, and LLaVA have further advanced instruction-following and multimodal generalization \cite{alayrac2022flamingo,li2023blip,liu2023visual}. Domain-specific models such as DeepSeek-VL have also emerged to address real-world vision--language understanding tasks \cite{lu2024deepseek}.

Within the building and energy domain, however, the use of LLMs to refine occupancy counts derived from surveillance video remains largely unexplored in systematic comparative studies. Existing research has primarily focused on detection models, sensor fusion, or predictive modeling, with limited benchmarking of LLM-based refinement against strong MOT baselines using frame-level ground truth \cite{wang2024computer}. This gap motivates the inclusion of an LLM-based refinement pipeline in the present study and its controlled comparison with detection-only and tracking-based approaches under identical experimental conditions.

In summary, existing literature establishes occupancy as a critical variable for OCC and highlights limitations in both non-vision sensing and detection-only vision-based approaches. While multi-object tracking improves temporal consistency, it remains sensitive to detection quality and association errors. At the same time, LLMs and VLMs offer a promising mechanism for reasoning-based refinement of perception outputs, yet their role in building occupancy measurement has not been rigorously explored. Building on these foundations, this work presents an experimental evaluation of detection-only, tracking-based, and LLM-refined occupancy measurement pipelines using real indoor surveillance video with manual ground-truth annotation.

\section{METHOD}
\label{section3}

\begin{figure*}[t]
    \centering
    \includegraphics[width=\textwidth]{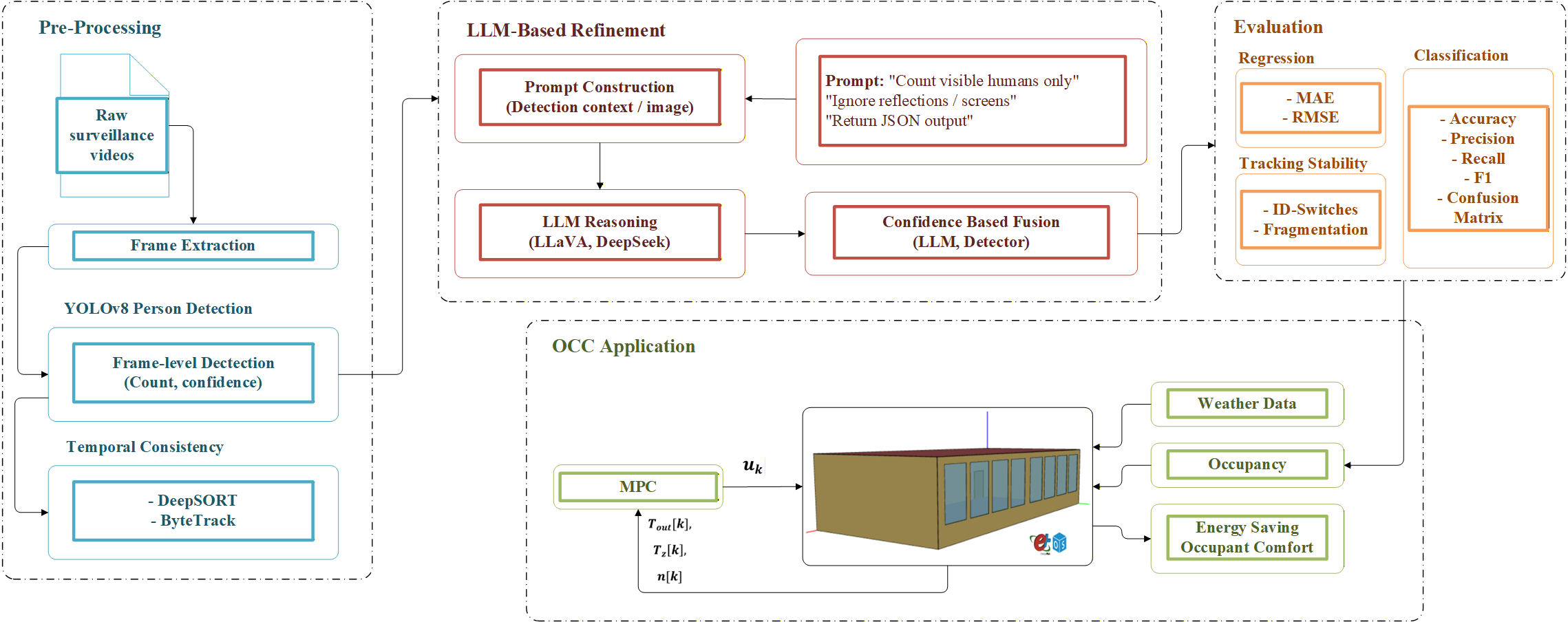}
    \caption{Overview of the proposed occupancy measurement and control integration framework.}
    \label{fig:method}
\end{figure*}

This section presents the methodological framework for surveillance-video-based indoor occupancy measurement in support of OCC. The study is fully experimental and focuses on occupancy measurement from surveillance video, rather than model training or occupancy prediction. Four perception pipelines are implemented and evaluated under identical input conditions, enabling a controlled comparison of detection-only, tracking-based, and reasoning-enhanced approaches. The resulting occupancy signals are structured to be directly compatible with downstream MPC-based HVAC control. The overall workflow is summarized in Fig.~\ref{fig:method}.

\subsection{Framework Overview}

As illustrated in Fig.~\ref{fig:method}, the proposed framework consists of four sequential stages: data preprocessing, occupancy measurement, performance evaluation, and OCC application. Raw surveillance videos are first converted into frame-level images through a standardized preprocessing pipeline. These frames are then processed using a shared YOLOv8 person detector, which provides a unified detection backbone across all evaluated pipelines.

Based on the YOLOv8 detection outputs, four occupancy measurement pipelines are implemented: (A) frame-level counting based solely on detections; (B) detection combined with DeepSORT for identity-consistent tracking; (C) detection integrated with ByteTrack as a state-of-the-art multi-object tracking baseline; and (D) detection followed by refinement using an LLM. Each pipeline produces frame-level occupancy estimates aligned with the original video timestamps. This unified temporal representation supports aggregation to fixed control intervals and direct interfacing with MPC-based HVAC controllers. Manual ground-truth occupancy annotations are used only for evaluation and are not accessible during inference, ensuring an unbiased comparison across pipelines.

\subsection{Preprocessing and Ground-Truth Annotation}

Raw surveillance videos are processed by extracting individual frames at a uniform temporal resolution. This procedure ensures consistent temporal alignment across all pipelines and enables direct, frame-level comparison between estimated occupancy and reference annotations. Frames affected by data corruption or incomplete information are excluded to maintain data quality.

Ground-truth occupancy is obtained through exhaustive manual annotation using a custom labeling tool. For each extracted frame, the ground-truth occupancy is defined as the number of occupants visibly present within the camera’s field of view. Annotation guidelines are applied consistently across all videos, including frames with partial occlusion, overlapping occupants, or limited visibility. To improve temporal coherence, annotations are reviewed across adjacent frames and corrected when inconsistencies are identified. The resulting frame-level ground truth serves as the reference for all quantitative evaluation metrics reported in this study.

\subsection{Occupancy Measurement Models}

YOLOv8 is adopted as the person detection backbone for all evaluated pipelines. For each input frame $t$, YOLOv8 outputs a set of bounding box detections corresponding to the person class, each associated with a confidence score. A fixed confidence threshold is applied to filter out low-confidence detections. In the detection-only pipeline, the frame-level occupancy estimate $\hat{y}_t$ is obtained by directly counting the retained detections.

To improve temporal consistency, YOLOv8 detections are further processed using multi-object tracking (MOT). In the DeepSORT-based pipeline, detections are associated across consecutive frames using a combination of motion modeling and appearance embeddings, resulting in identity-consistent trajectories. In the ByteTrack-based pipeline, both high- and low-confidence detections are incorporated during the association process, improving robustness to missed detections and reducing track fragmentation. For both tracking-based pipelines, the frame-level occupancy estimate is computed as the number of active track identities at frame $t$.

Although tracking-based methods reduce frame-to-frame fluctuations, their performance remains sensitive to detector quality and association reliability, particularly under occlusion and abrupt motion. To further address these limitations, an LLM-based refinement stage is introduced as a post-processing module.

\subsubsection{YOLOv8 + LLM-based Refinement and Fusion}
\label{subsec:yolo_llm}

An LLM-based refinement module is introduced to improve occupancy estimates in frames affected by ambiguity, such as occlusion, abrupt count changes, or low detector confidence. This module operates as a post-detection reasoning layer and does not replace the underlying detector or tracking components. YOLOv8 outputs are converted into structured representations that include frame identifiers, detector-derived counts, mean confidence scores, and short-term temporal context. Based on these inputs, the LLM performs conservative reasoning to identify implausible occupancy variations and to correct false positive or false negative estimates.

\paragraph{Frame selection for LLM review.}
After YOLOv8 inference, each frame $t$ produces (i) a detector-based count $c_t$ and (ii) a mean detection confidence $\bar{s}_t \in [0,1]$, computed over all retained person detections in that frame. A lightweight screening step selects frames for LLM review when detector uncertainty or potentially problematic conditions are observed. These include low confidence with nonzero counts, boundary cases such as $c_t \in \{0,1\}$, and unusually high counts relative to neighboring frames. Selected frames are grouped into small batches to limit the number of LLM calls.

\paragraph{LLM prompting and outputs.}
For each selected batch, the LLM is instructed to return strict JSON outputs containing, for every queried frame $t$, an integer corrected occupancy count $c^{\text{llm}}_t \ge 0$ and a corresponding confidence score $s^{\text{llm}}_t \in [0,1]$. The prompt enforces conservative behavior under uncertainty and disallows explanatory text, ensuring that outputs remain machine-parseable and suitable for automated fusion. Two LLM configurations are evaluated: (i) the vision--language model LLaVA~7B, which receives image inputs for visual reasoning, and (ii) the text-only large language model DeepSeek, which operates solely on structured detector outputs and temporal context.

\paragraph{LLM-based refinement and fusion.}
The refinement and fusion process is summarized in Algorithm~\ref{alg:llm_refinement}. For each reviewed frame, the LLM-refined count $c^{\text{llm}}_t$ replaces the detector-derived count $c_t$ only when the LLM confidence exceeds the detector confidence by a predefined margin. This confidence-gated fusion strategy limits unnecessary corrections and ensures that LLM intervention is applied only when sufficient contextual evidence is available. The resulting occupancy estimates are then used for subsequent evaluation and control-oriented analysis.

\begin{algorithm}[t]
\caption{LLM-Based Refinement via Prompted Reasoning}
\label{alg:llm_refinement}
\begin{algorithmic}[1]
\REQUIRE Detector-derived information for frame $t$:
$\{c_t, \bar{s}_t, c_{t-1}, c_{t+1}, r_t\}$
\ENSURE Refined occupancy estimate $\hat{y}_t$ and associated confidence $\hat{s}_t$

\STATE Construct a structured prompt $\mathcal{P}_t$ containing:
\STATE \quad (i) frame identifier and timestamp
\STATE \quad (ii) detector person count $c_t$ and mean confidence $\bar{s}_t$
\STATE \quad (iii) temporal context via neighboring counts $\{c_{t-1}, c_{t+1}\}$
\STATE \quad (iv) refinement rules:
\STATE \quad \quad • count only physically present humans
\STATE \quad \quad • prefer conservative lower counts under uncertainty
\STATE \quad \quad • return output strictly in machine-readable JSON format

\STATE Query the LLM with prompt $\mathcal{P}_t$ to perform reasoning-based refinement

\STATE Receive LLM output:
$\{c^{\mathrm{llm}}_t, s^{\mathrm{llm}}_t\}$

\IF{$s^{\mathrm{llm}}_t \ge \bar{s}_t + \delta$}
    \STATE $\hat{y}_t \leftarrow c^{\mathrm{llm}}_t$
    \STATE $\hat{s}_t \leftarrow s^{\mathrm{llm}}_t$
\ELSE
    \STATE $\hat{y}_t \leftarrow c_t$
    \STATE $\hat{s}_t \leftarrow \bar{s}_t$
\ENDIF

\RETURN $\hat{y}_t, \hat{s}_t$
\end{algorithmic}
\end{algorithm}

\paragraph{Vision--language refinement.}
In the vision--language configuration, each LLM call processes a small batch of image frames. The prompt specifies indoor occupancy counting rules, including ignoring individuals appearing on screens, posters, reflections, or outside the physical room boundary. For each queried frame $t$, the model returns a corrected occupancy count $c^{\mathrm{llm}}_t$ together with an associated confidence score $s^{\mathrm{llm}}_t$, both grounded directly in visual evidence.

\paragraph{Text-only refinement.}
In the text-only configuration, the LLM does not access raw images. Instead, each frame is represented by structured attributes including a frame identifier, timestamp (if available), detector count $c_t$, mean detector confidence $\bar{s}_t$, neighboring detector counts $(c_{t-1}, c_{t+1})$, and a short flag indicating the reason for LLM review. Using this temporal and contextual information, the LLM suppresses implausible fluctuations and corrects systematic detector errors without direct visual input.

\paragraph{Detector--LLM fusion rule.}
The final occupancy estimate $\tilde{c}_t$ is obtained using a conservative confidence-margin fusion rule. Let $\Delta > 0$ denote a fixed confidence margin (set to $\Delta = 0.15$). For each reviewed frame,
\begin{equation}
\tilde{c}_t =
\begin{cases}
c^{\mathrm{llm}}_t, & \text{if } s^{\mathrm{llm}}_t \ge \bar{s}_t + \Delta, \\
c_t, & \text{otherwise},
\end{cases}
\label{eq:fusion_rule}
\end{equation}
and if $(c^{\mathrm{llm}}_t, s^{\mathrm{llm}}_t)$ is unavailable due to parsing or inference failure, the detector output $c_t$ is retained. This strategy ensures that LLM-based corrections are applied only when the LLM expresses sufficiently higher confidence than the detector proxy.

\paragraph{Occupied/unoccupied state.}
For OCC-related evaluation, the fused occupancy count is mapped to a binary occupancy state:
\begin{equation}
\tilde{z}_t =
\begin{cases}
1, & \tilde{c}_t \ge 1, \\
0, & \tilde{c}_t = 0,
\end{cases}
\end{equation}
where $\tilde{z}_t = 1$ denotes an occupied frame.

\paragraph{Implementation notes.}
LLM outputs are stored on a per-video basis and merged with detector-derived results to form a validated occupancy time series. Failed LLM calls or malformed responses are logged, and the pipeline safely defaults to detector outputs, ensuring robustness and reproducibility during large-scale experimental evaluation.

\subsection{Occupant-Centric HVAC Control}

To demonstrate the practical usability of the proposed occupancy measurement framework for building operation, the estimated occupancy signals are integrated into an MPC-based supervisory HVAC control strategy. Occupancy estimates are treated as exogenous inputs to the control layer and are not used to train or adapt the controller.

\subsubsection{Model Predictive Control-Based OCC}

To examine how reliable occupancy measurements can support predictive HVAC operation, a simplified MPC-based OCC strategy is considered. In this study, MPC is applied at a supervisory level using its standard receding-horizon formulation, consistent with external control implementation through the EnergyPlus PythonPlugin interface, rather than emphasizing detailed objective function design.

At each control step $k$, corresponding to a discrete EnergyPlus simulation time step of 5 minutes, the controller predicts system behavior over a finite prediction horizon of $N$ future steps, where $N = 12$ represents one hour. Based on the predicted evolution of zone conditions and anticipated occupancy, the controller computes a sequence of future heating and cooling thermostat setpoints:
\begin{equation}
\{T_{\mathrm{htg}}[k], T_{\mathrm{clg}}[k], \ldots,
T_{\mathrm{htg}}[k+N-1], T_{\mathrm{clg}}[k+N-1]\},
\end{equation}
where $T_{\mathrm{htg}}[k]$ and $T_{\mathrm{clg}}[k]$ denote the heating and cooling setpoint temperatures (in~$^\circ$F) applied at control step $k$, respectively.

\paragraph{Prediction Model}
Zone thermal behavior is approximated using a lightweight discrete-time model:
\begin{equation}
T_z[k+1] = a T_z[k] + b T_{\mathrm{out}}[k] + c u[k] + d n[k],
\end{equation}
where $T_z[k]$ is the zone air temperature at step $k$, $T_{\mathrm{out}}[k]$ is the forecast outdoor air temperature obtained from the EnergyPlus weather file, $n[k]$ denotes the estimated zone occupancy count derived from the proposed perception pipelines, and $a$, $b$, $c$, and $d$ are scalar coefficients identified from simulation data. The term $u[k]$ represents the supervisory control input derived from thermostat setpoints.

For compact representation, the control input is defined using the midpoint setpoint:
\begin{equation}
T_{\mathrm{mid}}[k] =
\frac{T_{\mathrm{htg}}[k] + T_{\mathrm{clg}}[k]}{2},
\end{equation}
which serves as a practical proxy for HVAC actuation intensity within the EnergyPlus supervisory control framework.

\paragraph{Receding-Horizon Implementation}
At each control interval, the MPC solves the optimization problem over the prediction horizon and produces an optimal sequence of setpoints. Following the receding-horizon principle, only the first setpoint pair is applied to the building model:
\begin{equation}
\{T_{\mathrm{htg}}^{\ast}[k], T_{\mathrm{clg}}^{\ast}[k]\}.
\end{equation}
These values are written to EnergyPlus thermostat actuators via the PythonPlugin interface. At the subsequent control step, the time index advances,
\begin{equation}
k \leftarrow k+1,
\end{equation}
and the optimization is repeated using updated zone states, occupancy estimates, and weather forecasts.

This rolling optimization process enables closed-loop interaction between the occupancy-aware supervisory controller and the EnergyPlus simulation. By continuously re-optimizing based on updated occupancy information, the MPC-based OCC strategy can anticipate future space usage, initiate preconditioning ahead of occupancy, and relax conditioning during predicted low-occupancy periods.

Overall, this formulation demonstrates how accurate and temporally stable occupancy measurements can be directly embedded into predictive HVAC control logic within EnergyPlus, even when MPC is implemented in a simplified and structure-focused manner.

\subsection{Evaluation Metrics}

The performance of the proposed occupancy measurement pipelines is evaluated using occupancy accuracy metrics and tracking stability metrics. These metrics are selected to reflect the requirements of OCC, where reliable HVAC operation depends on both accurate numerical estimation and temporally stable occupancy signals, particularly when such signals are used as inputs to predictive control algorithms.

Let $y_t \in \mathbb{N}$ denote the ground-truth occupant count at frame $t$,
$\hat{y}_t \in \mathbb{N}$ the corresponding estimated count produced by a given pipeline,
and $T$ the total number of evaluated frames.

\subsubsection{Occupancy Accuracy Metrics}

The mean absolute error (MAE) measures the average absolute deviation between estimated and reference occupancy counts:
\begin{equation}
\mathrm{MAE} = \frac{1}{T} \sum_{t=1}^{T} \left| \hat{y}_t - y_t \right|.
\end{equation}

The root mean square error (RMSE) assigns greater weight to larger deviations:
\begin{equation}
\mathrm{RMSE} =
\sqrt{\frac{1}{T} \sum_{t=1}^{T} \left( \hat{y}_t - y_t \right)^2 }.
\end{equation}

Exact occupancy accuracy quantifies the proportion of frames for which the estimated count exactly matches the ground truth:
\begin{equation}
\mathrm{Accuracy}_{\mathrm{exact}} =
\frac{1}{T} \sum_{t=1}^{T} \mathbb{I}\!\left( \hat{y}_t = y_t \right),
\end{equation}
where $\mathbb{I}(\cdot)$ denotes the indicator function.

For OCC-related evaluation, occupancy is additionally represented as a binary state. The ground-truth and estimated occupancy states are defined as
\begin{equation}
z_t =
\begin{cases}
1, & y_t > 0, \\
0, & y_t = 0,
\end{cases}
\qquad
\hat{z}_t =
\begin{cases}
1, & \hat{y}_t > 0, \\
0, & \hat{y}_t = 0.
\end{cases}
\end{equation}

Based on these definitions, precision, recall, and F1-score are computed as
\begin{equation}
\mathrm{Precision} = \frac{TP}{TP + FP}, \qquad
\mathrm{Recall} = \frac{TP}{TP + FN},
\end{equation}
\begin{equation}
\mathrm{F1} =
2 \cdot \frac{\mathrm{Precision} \cdot \mathrm{Recall}}
{\mathrm{Precision} + \mathrm{Recall}},
\end{equation}
where $TP$, $FP$, and $FN$ denote true positives, false positives, and false negatives, respectively.

\subsubsection{Confusion Matrix Analysis}

Binary occupancy detection performance is further characterized using a confusion matrix:
\begin{equation}
\mathbf{C} =
\begin{bmatrix}
TN & FP \\
FN & TP
\end{bmatrix},
\end{equation}
where $TN$ represents correctly identified unoccupied frames,
$TP$ correctly identified occupied frames,
$FP$ false alarms during unoccupied periods,
and $FN$ missed detections when occupants are present.

From an OCC perspective, false negatives are particularly critical because undetected occupancy may result in insufficient ventilation or thermal conditioning, whereas false positives primarily lead to unnecessary energy use. The confusion matrix therefore provides complementary diagnostic insight beyond aggregated accuracy metrics.

\subsubsection{Tracking Stability Metrics}

For pipelines incorporating multi-object tracking, temporal consistency is evaluated using identity-based metrics. An identity switch (ID-switch) is recorded when a tracked individual is assigned different identities across successive frames, indicating association inconsistency.

Fragmentation measures the extent to which a single ground-truth trajectory is divided into multiple track segments over time. Higher fragmentation reflects reduced temporal continuity and may introduce artificial fluctuations in identity-based occupancy estimates.

Collectively, these metrics enable a comprehensive assessment of numerical accuracy, robustness of occupancy state detection, and temporal stability across detection-only, tracking-based, and LLM-refined occupancy measurement pipelines.

\section{EXPERIMENT}
\label{section4}

This section describes the experimental environment, dataset, and evaluation protocol used to assess the proposed surveillance-video-based occupancy measurement pipelines. Experiments are conducted under realistic indoor conditions in an operational research laboratory, with the objective of evaluating occupancy accuracy and temporal stability in a setting representative of occupant-centric control (OCC) deployment. In addition, the experimental results are used to examine how occupancy measurement quality propagates to downstream HVAC control performance under an MPC-based supervisory control framework.

\begin{figure*}[t]
    \centering
    \includegraphics[width=\textwidth]{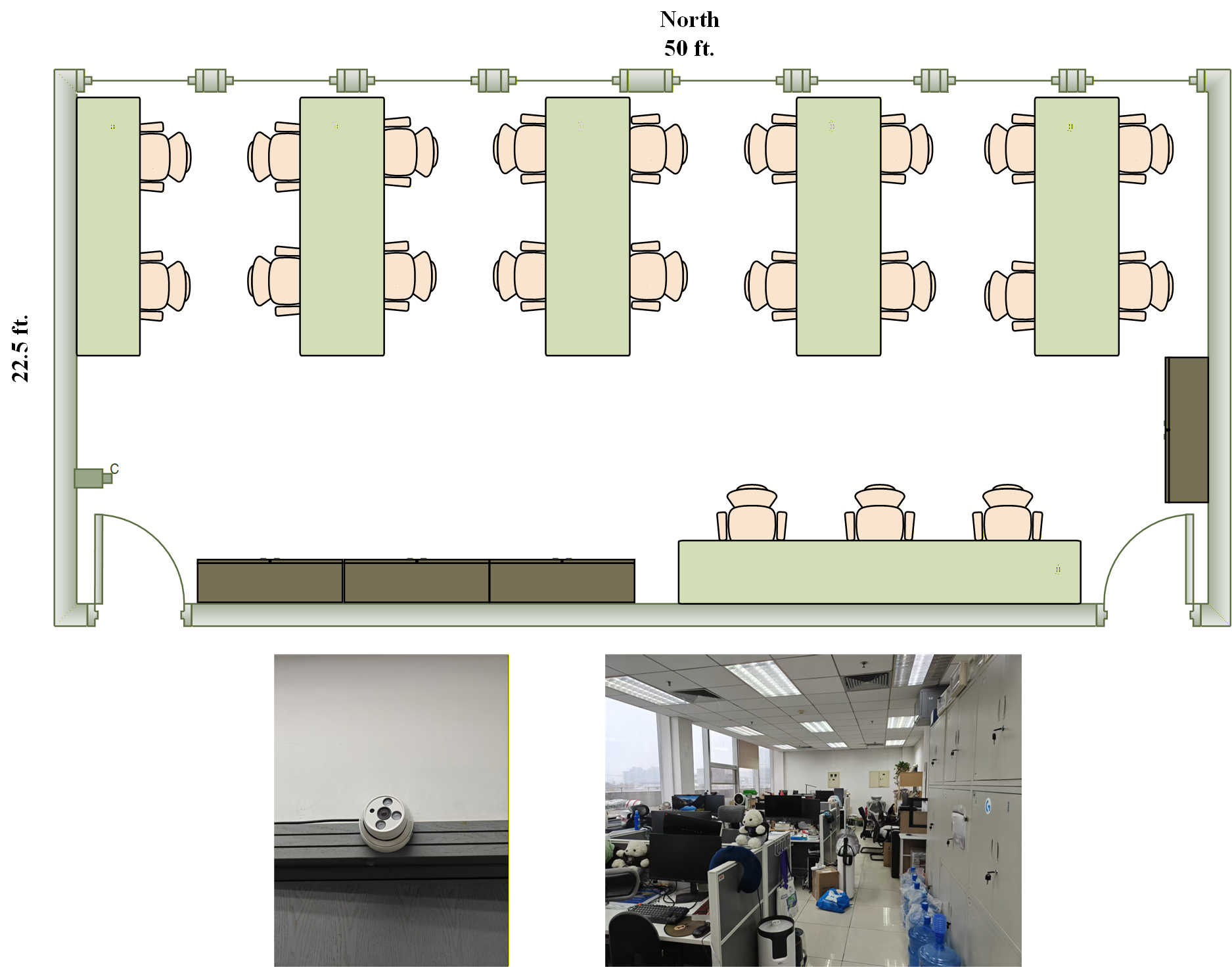}
    \caption{Floor plan and representative views of the experimental laboratory environment at FIT, Tsinghua University.}
    \label{fig:floor}
\end{figure*}

\subsection{Experimental Environment and Dataset}

The experimental site is a PhD research laboratory located at the Future Internet Technology Research Center (FIT), Tsinghua University, Beijing, as shown in Fig.~\ref{fig:floor}. The space is an open-plan indoor laboratory with multiple workstations, desks, and shared areas. Occupancy varies over time due to arrivals, departures, and intermittent activities. A fixed surveillance camera is installed to monitor the space, capturing common indoor challenges such as partial occlusion, overlapping occupants, and variations in lighting conditions.

Surveillance video data are collected over five non-consecutive days: 30 October, 2 November, 6 November, 12 November, and 18 November. Recordings span different times of day, resulting in diverse occupancy patterns and scene dynamics. Individual video clips range in duration from approximately 6--10 seconds to approximately 2--2.5 minutes, depending on the recording session and observed activity.

In total, the dataset comprises 1,188 video clips and 20,722 extracted frames. All videos are processed using a uniform preprocessing procedure to generate frame-level images, which serve as inputs to the occupancy measurement pipelines. For control-oriented evaluation, frame-level occupancy estimates are subsequently aggregated to 5-minute intervals to match the supervisory control and MPC time-step requirements in EnergyPlus.

\subsection{Ground Truth and Experimental Protocol}

Frame-level ground-truth occupancy is obtained through manual inspection of all extracted frames. For each frame, the number of occupants visibly present within the camera’s field of view is recorded using a custom annotation tool. Consistent labeling guidelines are applied across the entire dataset, including frames with partial visibility, overlapping occupants, or occlusion. This exhaustive annotation process yields a complete and frame-aligned ground-truth occupancy record for every frame in the dataset.

All frames are processed independently by the four occupancy measurement pipelines described in Section~\ref{section3}: (A) YOLOv8-only frame-level detection, (B) YOLOv8 combined with DeepSORT tracking, (C) YOLOv8 combined with ByteTrack tracking, and (D) YOLOv8 with LLM-based refinement. A single, fixed YOLOv8 detector configuration is used across all pipelines to ensure methodological consistency and fair comparison.

Ground-truth annotations are not accessible during inference and are used exclusively for post hoc evaluation. No video-specific parameter tuning, retraining, or adaptive configuration is performed. Each pipeline produces a frame-level occupancy time series aligned with the original video timestamps, enabling direct and unbiased comparison of performance across methods. These occupancy time series also serve as exogenous inputs for subsequent MPC-based HVAC simulations.

\subsection{Evaluation Procedure and Scope}

All evaluation metrics defined in Section~\ref{section3} are computed at the frame level by comparing estimated occupancy counts with the corresponding manual ground truth. For pipelines incorporating multi-object tracking, identity-based stability metrics, including ID-switches and fragmentation, are derived directly from the tracker outputs. Performance is first summarized at the level of individual video clips and subsequently aggregated across the full dataset to obtain overall results.

This aggregation strategy reflects practical deployment requirements, where occupancy measurement systems must operate reliably across diverse temporal segments, activity patterns, and occupancy conditions, rather than being optimized for a limited subset of scenes. In addition to perception-level evaluation, aggregated occupancy estimates are used to drive MPC-based HVAC control simulations in OpenStudio--EnergyPlus over a one-year period, enabling quantitative assessment of how occupancy measurement quality affects energy consumption and thermal comfort.

The experimental evaluation focuses primarily on the numerical accuracy and temporal stability of surveillance-video-based occupancy measurement. The HVAC control results are intended to illustrate downstream implications of sensing quality rather than to provide a comprehensive control-theoretic optimization study. It is assumed that the camera field of view provides sufficient coverage of the monitored space and that observed occupancy is representative of zone-level occupancy for benchmarking purposes within an OCC context.

\section{RESULTS}
\label{section5}

This section reports the experimental results of the surveillance-video-based occupancy measurement pipelines. Performance is evaluated in terms of counting accuracy, occupied/unoccupied classification reliability, and temporal stability, with emphasis on implications for occupant-centric control (OCC). Global results aggregated across all videos and frames are summarized in Tables~\ref{tab:global_metrics_grouped}--\ref{tab:tracking_metrics}.

\begin{table*}[t]
\centering
\caption{Global evaluation metrics across models.}
\label{tab:global_metrics_grouped}
\resizebox{\linewidth}{!}{%
\begin{tabular}{l r rr rrrrrr}
\hline
 &  & \multicolumn{2}{c}{\textbf{Counting (Regression)}} & \multicolumn{6}{c}{\textbf{Occupied/Unoccupied (Classification)}} \\
\cline{3-4}\cline{5-10}
\textbf{Model} & $\mathbf{n_{\text{frames}}}$ & \textbf{MAE} & \textbf{RMSE} & \textbf{Accuracy} & \textbf{Precision} & \textbf{Recall} & \textbf{F1} & \textbf{FN} & \textbf{FP} \\
\hline
YOLOv8-only   & 20722 & 1.0984 & 1.5827 & 0.8125 & 0.8796 & 0.9051 & 0.8922 & 1685 & 2201 \\
YOLOv8 +DeepSORT & 20722 & 1.1653 & 1.6348 & 0.7964 & 0.8728 & 0.8926 & 0.8826 & 1907 & 2311 \\
YOLOv8 +ByteTrack & 20722 & 1.0502 & 1.5419 & 0.8193 & 0.8789 & 0.9154 & 0.8962 & 1503 & 2241 \\
YOLOv8 +VLM (Ollama/LLaVA) & 20722 & 0.9648 & 1.4216 & 0.8525 & 0.9021 & 0.9287 & 0.9152 & 1266 & 1790 \\
\textbf{YOLOv8 +LLM (DeepSeek)} & 20722 & \textbf{0.8923} & \textbf{1.3124} & \textbf{0.8824} & \textbf{0.9245} & \textbf{0.9396} & \textbf{0.9320} & \textbf{1073} & \textbf{1363} \\
\hline
\end{tabular}%
}
\end{table*}

\begin{table*}[t]
\centering
\caption{Confusion matrices for occupied/unoccupied classification (Occupied = positive class). Totals satisfy $(\mathrm{TN}+\mathrm{FP})=2961$ and $(\mathrm{FN}+\mathrm{TP})=17761$.}
\label{tab:confusion_matrices_pretty}
\resizebox{\linewidth}{!}{%
\begin{tabular}{l rr rr}
\toprule
\multirow{2}{*}{\textbf{Model}} &
\multicolumn{2}{c}{\textbf{Actual: Unoccupied}} &
\multicolumn{2}{c}{\textbf{Actual: Occupied}} \\
\cmidrule(lr){2-3}\cmidrule(lr){4-5}
& \textbf{Pred: Unocc. (TN)} & \textbf{Pred: Occ. (FP)} & \textbf{Pred: Unocc. (FN)} & \textbf{Pred: Occ. (TP)} \\
\midrule
YOLOv8-only               &  760 & 2201 & 1685 & 16076 \\
YOLOv8 +DeepSORT          &  650 & 2311 & 1907 & 15854 \\
YOLOv8 +ByteTrack         &  720 & 2241 & 1503 & 16258 \\
YOLOv8 +VLM (Ollama/LLaVA)& 1171 & 1790 & 1266 & 16495 \\
\textbf{YOLOv8 +LLM (DeepSeek)} & \textbf{1598} & \textbf{1363} & \textbf{1073} & \textbf{16688} \\
\bottomrule
\end{tabular}%
}
\end{table*}

\begin{table}[t]
\centering
\caption{Tracking stability metrics for identity-based occupancy measurement pipelines.}
\label{tab:tracking_metrics}
\resizebox{\linewidth}{!}{%
\begin{tabular}{lcc}
\toprule
\textbf{Model} & \textbf{ID-switches} & \textbf{Fragmentation} \\
\midrule
YOLOv8-only & -- & -- \\
YOLOv8 + DeepSORT & 2143 & 1687 \\
YOLOv8 + ByteTrack & 1326 & 1048 \\
YOLOv8 +VLLM (Ollama/LLaVA) & -- & -- \\
YOLOv8 + LLM (DeepSeek) & -- & -- \\
\bottomrule
\end{tabular}%
}
\end{table}

\subsection{Overall Occupancy Counting Performance}

Table~\ref{tab:global_metrics_grouped} summarizes global counting accuracy across all 20{,}722 evaluated frames. Among the detection-only and tracking-based baselines, YOLOv8+ByteTrack achieves the lowest counting error, outperforming YOLOv8-only and YOLOv8+DeepSORT, indicating that multi-object tracking improves temporal consistency and reduces frame-level counting fluctuations relative to detection alone.

Both LLM-refined pipelines further reduce counting error. YOLOv8+VLM (Ollama/LLaVA) improves robustness to systematic detection errors, while YOLOv8+LLM (DeepSeek) delivers the best overall counting performance. Relative to the best tracking baseline (YOLOv8+ByteTrack), the DeepSeek-refined pipeline shows lower MAE and RMSE, indicating that reasoning-based refinement corrects residual errors that persist after identity-consistent tracking.

These results suggest that LLM-based refinement addresses errors that are not fully resolved by tracking, including missed detections under occlusion and unstable counting results. YOLOv8+LLM (DeepSeek) also outperforms YOLOv8+VLM (Ollama/LLaVA). DeepSeek refines detector outputs using structured contextual information, which improves temporal consistency in the present fixed-camera indoor setting. This advantage is more evident when occupancy variation is limited and detection errors are repeated over time.

\subsection{Occupied/Unoccupied Classification Performance}

In OCC applications, false unoccupied decisions are critical as they may lead to inadequate ventilation or thermal discomfort. Accordingly, Table~\ref{tab:global_metrics_grouped} reports binary occupied/unoccupied classification metrics.

Detection-only and tracking-based pipelines provide moderate performance, with YOLOv8+ByteTrack offering the best baseline due to better recall for occupied frames. However, all baseline methods show false negatives and false positives, reflecting sensitivity to transient detection and association errors.

LLM-based refinement substantially improves classification metrics. YOLOv8+LLM (DeepSeek) achieves the highest accuracy, precision, recall, and F1-score, while also minimizing false negatives. This reduction is crucial for OCC, as missed occupancy increases the risk of comfort degradation and insufficient conditioning.

The improved classification performance of DeepSeek compared to LLaVA is due to the interaction between the refinement design and the experimental setup. DeepSeek uses structured contextual information to refine detector outputs, enhancing temporal consistency and occupancy decision quality. In the fixed-view laboratory environment, with limited occupancy transitions, this approach outperforms image-level reasoning.

The confusion matrices in Table~\ref{tab:confusion_matrices_pretty} highlight these trends. Compared with baselines, the DeepSeek-refined pipeline increases true positives and true negatives while reducing both false positives and false negatives. This balanced improvement shows that reasoning-based refinement enhances decision consistency, rather than merely biasing predictions toward the occupied class. The relative performance of DeepSeek and LLaVA depends on deployment conditions; in larger or more visually heterogeneous spaces with greater occupancy ranges and perspective variation, direct visual reasoning may offer greater benefits than in the present small-scale laboratory setup.

\subsection{Tracking Stability Analysis}

Temporal stability is evaluated using identity-based metrics for the tracking pipelines, as reported in Table~\ref{tab:tracking_metrics}. YOLOv8+ByteTrack produces substantially fewer ID-switches and lower fragmentation than YOLOv8+DeepSORT, confirming superior association robustness under realistic indoor conditions.

The LLM-based pipelines do not generate explicit identity trajectories and are therefore not evaluated using tracking-specific metrics. However, their improved counting and classification performance suggests that reasoning-based refinement can mitigate downstream effects of identity instability without relying on persistent track identities. This highlights a key distinction between identity-based temporal stabilization and reasoning-based occupancy correction, with the latter providing a complementary pathway for generating control-relevant occupancy signals.

\subsection{Implications for Occupant-Centric Control}

\begin{table*}[t]
\centering
\caption{Simulation and MPC supervisory control parameters.}
\label{tab:mpc_parameters}
\resizebox{\linewidth}{!}{%
\begin{tabular}{ll}
\toprule
\textbf{Item} & \textbf{Value} \\
\midrule
Weather file & Beijing, China (EnergyPlus EPW) \\
Simulation time step & 5 minutes \\
Simulation horizon & Annual (extrapolated from measured days) \\
Comfort model & Fanger PMV/PPD \\
\midrule
Occupancy aggregation & Frame-level $\rightarrow$ 5-minute intervals \\
Occupied state & Occupied if count $\ge 1$ \\
Occupancy decision rule & Occupied if aggregated count $\ge 1$; unoccupied if count $=0$ \\
\midrule
Control strategy & Supervisory MPC (PythonPlugin) \\
Prediction horizon $N$ & 12 steps (1 hour) \\
Control update rate & 5 minutes \\
Control variables & Heating and cooling setpoints \\
Heating bounds & $60^\circ$F -- $70^\circ$F \\
Cooling bounds & $75^\circ$F -- $80^\circ$F \\
Setback logic (unoccupied) & Allow lower heating setpoint / higher cooling setpoint within MPC bounds \\
Occupied operation & MPC selects setpoints within bounds for comfort--energy trade-off \\
Setpoint values used & Dynamic (optimized) within bounds; not fixed single occupied/unoccupied values \\
Actuator interface & EnergyPlus thermostat actuators \\
\bottomrule
\end{tabular}%
}
\end{table*}

\begin{figure*}[h]
    \centering
    \includegraphics[width=\textwidth]{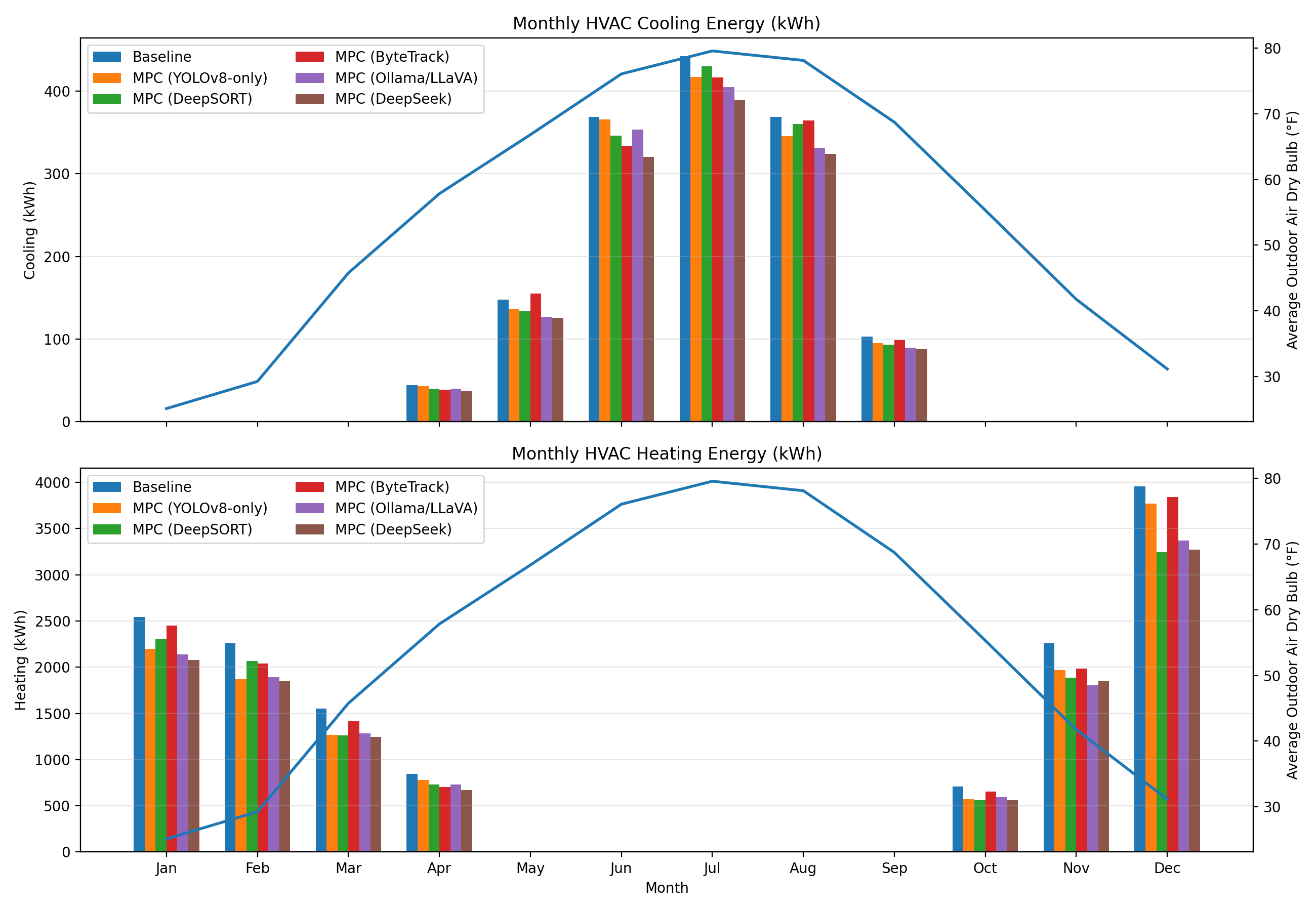}
    \caption{Monthly HVAC cooling and heating energy consumption under baseline and OCC operation.}
    \label{fig:load}
\end{figure*}

\begin{figure*}[h]
    \centering
    \includegraphics[width=\textwidth]{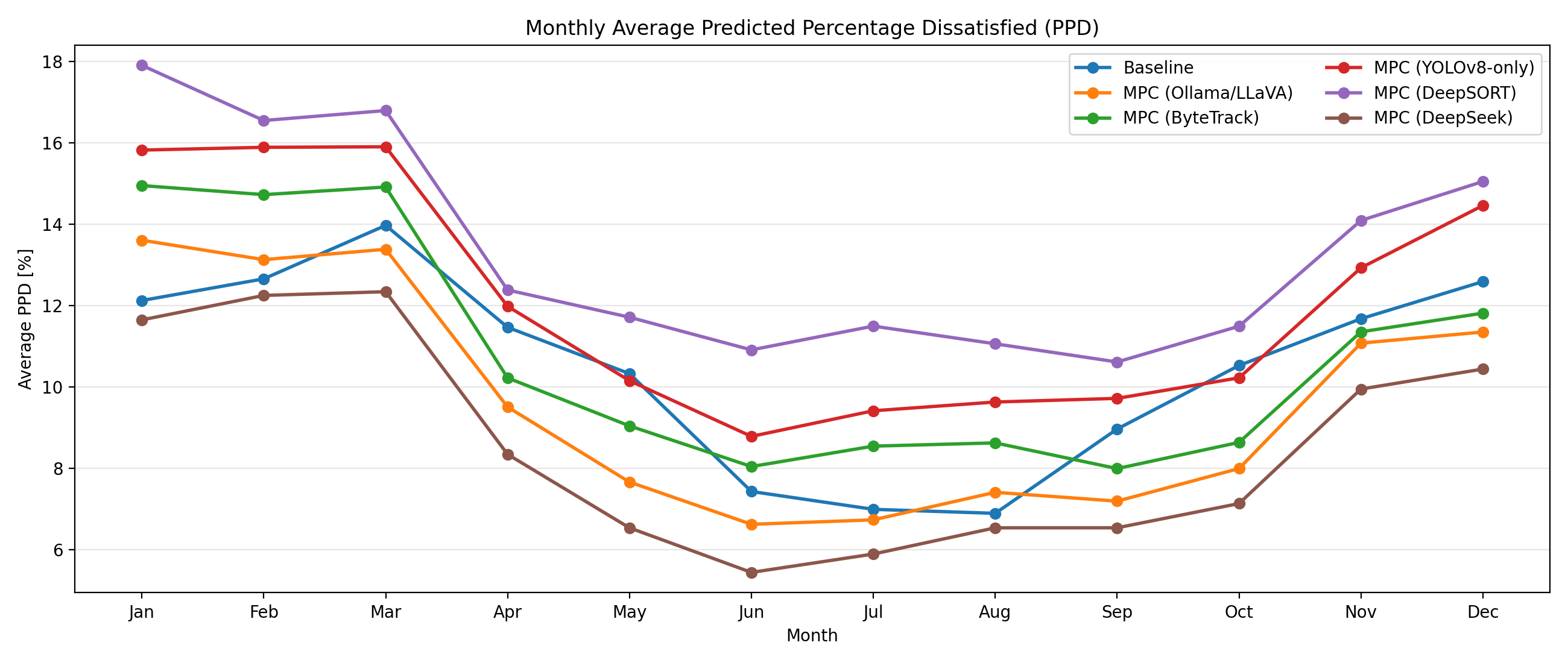}
    \caption{Monthly average predicted percentage dissatisfied (PPD) under baseline and OCC operation.}
    \label{fig:PPD}
\end{figure*}

\begin{table*}[t]
\centering
\caption{Annual average predicted percentage dissatisfied (PPD), HVAC energy consumption, and savings relative to the baseline.}
\label{tab:annual_energy_comfort_summary}
\resizebox{\linewidth}{!}{%
\begin{tabular}{l r rrrrrr}
\toprule
\multirow{2}{*}{Scenario} &
\multirow{2}{*}{Avg. PPD (\%)} &
\multicolumn{3}{c}{Annual Energy Consumption (kWh)} &
\multicolumn{3}{c}{Savings vs. Baseline (\%)} \\
\cmidrule(lr){3-5} \cmidrule(lr){6-8}
& & Cooling & Heating & Total HVAC & Cooling & Heating & Total HVAC \\
\midrule
Baseline           & 10.4699 & 1475.1 & 14127.8 & 15602.9 & 0.00  & 0.00  & 0.00  \\
MPC (YOLOv8-only)  & 12.0762 & 1403.1 & 12421.5 & 13824.6 & 4.88  & 12.08 & 11.40 \\
MPC (DeepSORT)     & 13.3376 & 1403.3 & 12047.9 & 13451.2 & 4.87  & 14.72 & 13.79 \\
MPC (ByteTrack)    & 10.7403 & 1408.1 & 13089.0 & 14497.1 & 4.54  & 7.35  & 7.09  \\
MPC (Ollama/LLaVA) & 9.6414  & 1346.4 & 11811.5 & 13157.9 & 8.72  & 16.40 & 15.67 \\
\textbf{MPC (DeepSeek)} & \textbf{8.5905} & \textbf{1284.0} & \textbf{11519.2} & \textbf{12803.2} & \textbf{12.96} & \textbf{18.46} & \textbf{17.94} \\
\bottomrule
\end{tabular}%
}
\end{table*}

To quantify how occupancy measurement quality propagates to building operation, each perception pipeline is coupled with the same MPC-based OCC supervisory controller and evaluated against a schedule-based baseline using OpenStudio--EnergyPlus simulations. Occupancy estimates from all pipelines are aggregated to 5-minute intervals to match the simulation time step. All simulation and control settings are kept identical across cases (Table~\ref{tab:mpc_parameters}), including the occupancy decision rule, setback logic, thermostat bounds, prediction horizon, update rate, and actuator interface. Therefore, differences in HVAC energy use and PPD are attributed primarily to the accuracy and temporal reliability of the occupancy input signals.

For all OCC cases, the same occupancy-driven setback policy is applied: a 5-minute interval is classified as occupied when the aggregated count is $\geq 1$, and unoccupied when the count is $0$. The MPC optimizes heating and cooling setpoints within the fixed bounds in Table~\ref{tab:mpc_parameters}. During unoccupied intervals, the optimizer may move setpoints toward energy-saving values (lower heating setpoint and/or higher cooling setpoint), whereas during occupied intervals it selects comfort--energy trade-off setpoints within the same bounds. The controller is implemented in receding-horizon form, updates every 5 minutes, and applies only the first optimized setpoint pair at each step. The reported heating and cooling values are allowable MPC setpoint ranges, not fixed occupied/unoccupied setpoints.

The experimental space is a PhD research laboratory with nominal 24/7 access. Manual inspection of ground-truth annotations confirms that the space is consistently unoccupied between 01{:}00 and 06{:}00. In the baseline scenario, HVAC operation follows a fixed dual-setpoint schedule that does not respond to actual occupancy variations.

Although occupancy measurements are obtained from five non-consecutive monitored days, these data are embedded into a full-year EnergyPlus simulation to evaluate long-term HVAC performance. Specifically, the measured occupancy profiles are reused as representative operational patterns within the annual simulation horizon, allowing annual energy and comfort impacts to be assessed under consistent weather and system assumptions. The selected setpoint bounds and setback behavior are used as a practical commercial-building-style OCC policy for comparative evaluation, rather than as universal optimal settings.

Figures~\ref{fig:load} and~\ref{fig:PPD}, together with Table~\ref{tab:annual_energy_comfort_summary}, summarize the resulting energy and comfort performance. Relative to the baseline total HVAC energy consumption of 15{,}602.9~kWh, all MPC-driven cases reduce HVAC energy use, with total savings ranging from 7.09\% to 17.94\%. The largest reduction is achieved by MPC driven by YOLOv8+LLM (DeepSeek), which yields 12{,}803.2~kWh total HVAC energy consumption (17.94\% savings).

The monthly PPD curves in Fig.~\ref{fig:PPD} provide a consistent picture of comfort performance throughout the year. DeepSeek maintains the lowest or near-lowest PPD in most months, while DeepSORT and YOLOv8-only generally exhibit higher PPD, especially during the winter and shoulder-season periods. This indicates that more reliable occupancy signals improve not only annual average comfort but also month-to-month control consistency.

These differences are consistent with the occupancy error characteristics in Tables~\ref{tab:global_metrics_grouped} and~\ref{tab:confusion_matrices_pretty}. False negative occupancy errors (occupied periods classified as unoccupied) are particularly harmful for OCC because they trigger setback during actual occupancy, increasing discomfort risk. Detector-only and tracking-based pipelines show relatively high false negative counts (e.g., 1{,}685 for YOLOv8-only and 1{,}907 for YOLOv8+DeepSORT), which is reflected in their higher average PPD values (12.08\% and 13.34\%, respectively). False positives mainly reduce energy savings by causing unnecessary conditioning during unoccupied periods.

Reasoning-enhanced pipelines mitigate both error modes. In particular, YOLOv8+LLM (DeepSeek) achieves the lowest false negative count (1{,}073) and a substantially lower false positive count (1{,}363), which translates into the best OCC performance: the highest energy savings and the lowest average PPD (8.59\%) in Table~\ref{tab:annual_energy_comfort_summary}. Compared with the baseline PPD of 10.47\%, this corresponds to an absolute reduction of 1.88 percentage points.

The better OCC performance of DeepSeek relative to LLaVA is related to the present experimental setting. The monitored space is a small, fixed-view laboratory with limited occupancy variation. In this setting, improvements in short-term consistency and reduction of false negatives provide larger downstream control benefits than additional image-level reasoning. This result may differ in larger or more visually heterogeneous spaces.

Overall, these results show that surveillance-video-based occupancy sensing is a control-critical component of OCC rather than only a perception task. Detector-only and tracking-based pipelines can reduce HVAC energy use, but unstable occupied/unoccupied decisions may degrade comfort. By contrast, reasoning-based refinement produces more control-suitable occupancy signals, enabling MPC-based OCC to improve both energy efficiency and predicted thermal comfort performance (lower PPD).

\subsection{Limitations and Future Work}

While this study demonstrates the effectiveness of surveillance-video-based occupancy measurement for OCC under realistic indoor conditions, several limitations remain. The evaluation is conducted in a single open-plan research laboratory with a fixed camera viewpoint. Further validation across a broader range of space types, such as classrooms, partitioned offices, and circulation areas, as well as different camera placements, is needed to assess generalizability.

The proposed pipelines rely exclusively on visual sensing. Although reasoning-based refinement improves robustness, performance may still degrade under severe occlusion, extreme crowding, or camera blind spots. Future work will explore multimodal extensions that integrate vision with privacy-preserving non-visual sensing to enhance reliability under challenging conditions. In addition, the MPC-based OCC implementation adopts a simplified supervisory architecture, emphasizing integration and control feasibility rather than full optimization. Future research will investigate tighter coupling between occupancy sensing and control, including uncertainty-aware MPC formulations, adaptive horizons, and learning-based controllers that explicitly account for sensing confidence.

\subsection{GenAI Disclosure}

The authors declare that artificial intelligence (AI) tools, specifically Gemini, were used solely to assist with language polishing and grammar checking of the manuscript text. All intellectual content, data analysis, interpretations, and conclusions were conceived, written, and verified by the authors.

\section{CONCLUSIONS}

This study presented a fully experimental evaluation of surveillance-video-based indoor occupancy measurement for occupant-centric control (OCC), benchmarking detection-only, tracking-based, and reasoning-enhanced perception pipelines using real indoor surveillance data with exhaustive frame-level ground truth. The results show that while multi-object tracking improves temporal stability relative to frame-level detection, it does not fully eliminate counting errors or misclassification of occupied states. Reasoning-based refinement using large language models further reduces counting error and, critically, suppresses false-negative occupancy decisions that are particularly detrimental for HVAC operation. Among all evaluated approaches, the YOLOv8+LLM (DeepSeek) pipeline achieves the most favorable balance between numerical accuracy, occupied/unoccupied classification reliability, and temporal robustness.

By integrating each occupancy measurement pipeline into the same MPC-based HVAC supervisory controller using OpenStudio--EnergyPlus, this work demonstrates that sensing quality directly propagates to control-level outcomes. Pipelines with elevated false-negative rates frequently misclassify occupied periods as unoccupied, causing the MPC controller to apply temperature setbacks during actual occupancy and leading to degraded thermal comfort despite moderate energy savings. In contrast, the reasoning-enhanced pipeline enables reliable setback during truly unoccupied periods and timely conditioning when occupancy is present or imminent, achieving the largest HVAC energy reduction (17.94\%) while also delivering the highest average thermal satisfaction (91.41\%). Overall, these findings confirm that occupancy sensing is a control-critical input rather than a standalone perception task, and that combining computer vision with reasoning-based refinement provides a practical and scalable pathway toward reliable, demand-responsive HVAC control in smart buildings.

\section{ACKNOWLEDGMENTS}

This work is supported by the National Natural Science Foundation of China under Grant No. 62192751, in part by Key R\&D Project of China under Grant No. 2017YFC0704100, the 111 International Collaboration Program of China under Grant No.B25027, and in part by the BNRist Program under Grant No. BNR2019TD01009, the National Innovation Center of High Speed Train R\&D project (CX/KJ-2020-0006).

\clearpage

\bibliographystyle{unsrt}

\bibliography{cas-refs}

\clearpage

\bio{}

\endbio

\end{document}